\documentclass[final,english]{bullsrsl}

\usepackage[latin1]{inputenc}
\usepackage[T1]{fontenc}

\usepackage{natbib} 

\usepackage{float}

\usepackage{graphicx}

\begin{document}
\title{Detection and identification of asteroids with the 4-m ILMT}

  \author[affil={1}, corresponding]{Sara}{Filali}
  \author[affil={2}]{Bhavya}{Ailawadhi}
  \author[affil={3,4}]{Talat}{Akhunov}
  \author[affil={2,5}]{Monalisa}{Dubey}
  \author[affil={2,5}]{Naveen}{Dukiya}
  \author[affil={6,7}]{Paul}{Hickson}
  \author[affil={2}]{Priyanshi}{Kumari}
  \author[affil={2}]{Gokul Singh}{Mehra}
  \author[affil={2}]{Kuntal}{Misra}
  \author[affil={8}]{Vibhore}{Negi}
  \author[affil={1}]{Anna}{Pospieszalska-Surdej}
  \author[affil={2,9}]{Kumar}{Pranshu}
  \author[affil={1,2}]{Jean}{Surdej}
  \author[affil={2,9}]{Sarvesh Kumar}{Yadav}

\affiliation[1]{Institute of Astrophysics and Geophysics, Liège University, Allée du 6 Août 19c, 4000 Liège, Belgium}
\affiliation[2]{Aryabhatta Research Institute of Observational Sciences, Manora Peak, Nainital, 263001 Uttarakhand, India}
\affiliation[3]{National University of Uzbekistan, Department of Astronomy and Astrophysics, 100174 Tashkent, Uzbekistan}
\affiliation[4]{Ulugh Beg Astronomical Institute of the Uzbek Academy of Sciences, Astronomicheskaya 33, 100052 Tashkent, Uzbekistan}
\affiliation[5]{Mahatma Jyotiba Phule Rohilkhand University, Pilibhit Bypass Road, Bareilly 243006, Uttar Pradesh, India}
\affiliation[6]{Department of Physics and Astronomy, The University of British Columbia, 6224 Agricultural Road, Vancouver, BC V6T 1Z1, Canada}
\affiliation[7]{Outer Space Institute, The University of British Columbia, 325-6224 Agricultural Road, Vancouver, BC V6T 1Z1, Canada}
\affiliation[8]{Kavli Institute for Astronomy and Astrophysics, Peking University, Beijing 100871, China}
\affiliation[9]{University of Calcutta, 87/1 College Street, Kolkata 700073, India}

\correspondance{sfilali@uliege.be}

\date{16th February 2026}

\maketitle

\begin{abstract}
The International Liquid Mirror Telescope (ILMT) covers a $22.3'$ wide strip of sky in declination ($\delta$), centred at $\delta = +29^\circ21'41.4''$ and right ascension ($\alpha$) in the range $0\,\mathrm{h} \le \alpha < 24\,\mathrm{h}$. Having a short focal length ($f/D \approx 2.4$) and a large diameter (4 m), makes the ILMT an excellent asteroid hunter.
The ILMT began its 4th cycle in October 2024, running through May 2025. The astrometric accuracy has been improved to $0.1''$, and the \texttt{PyLMT} --- a detection and classification pipeline ---, has been fine-tuned using data from previous cycles. The current detection rate is tens of transients detected each night with high accuracy in classification and identification.
We present statistical results for the asteroids detected during ILMT's Cycles 1-4. We first evaluate the astrometric performance of the detections across different ecliptic latitude ranges. We then describe the positions, apparent motions, and V magnitudes predicted by the Minor Planet Center (MPC) for the asteroids observed in the SDSS \textit{g'}, \textit{r'}, and \textit{i'} bands. Finally, we assess the ILMT's potential for detecting near-Earth objects (NEOs), potentially hazardous asteroids (PHAs), and comets.
\end{abstract}


\section{Introduction}

The 4-m International Liquid Miror Telescope (ILMT) located at the Devasthal observatory in the foothills of the Himalayas at $2378$ m altitude in India operates in a fixed zenithal configuration, repeatedly observing the same strip of sky as the Earth rotates. Owing to the site's latitude and longitude ($29^\circ 21'41.4''$, $79^\circ 41' 07.08''$), this observing mode provides a stable and highly reproducible survey footprint from year to year. Such a configuration is particularly well suited for the detection of transients, as it combines homogeneous image quality, minimal airmass variations, and recurrent coverage of identical sky regions, enabling both discovery and temporal linking of minor planets across multiple observing seasons.

Minor planets constitute a fundamental tracer of the formation and dynamical evolution of the Solar System, while near-Earth objects (NEOs) and potentially hazardous asteroids (PHAs) are of direct relevance for planetary defense. Efficient and automated identification of these bodies in wide-field time-domain data is therefore of major interest. Additionally, because ILMT's survey repeatedly scans the same celestial strip, it can provide a long-term temporal baseline for objects that re-enter the field, thereby improving orbit determination over multi-year timescales.

In this work we exploit this observing strategy to conduct a systematic search for minor planets in data obtained over four observing cycles. Source detection and classification are performed using \texttt{PyLMT}, an automated pipeline initially developed for transient detection \citep{pranshu2025pylmt}. During the first observing cycle, \texttt{PyLMT} has demonstrated a strong capability to identify asteroids \citep{filali2026asteros}. It was subsequently trained and fine-tuned using the first-cycle data, and is here applied uniformly to data from Cycles 1-4 to assess its performance and scientific return on an extended dataset.

\section{Observations and survey configuration}

In the ILMT's drift-scan configuration, the effective integration time along right ascension is 102.36 s \citep{surdej20254}. This short exposure limits trailing of moving objects while keeping sufficient sensitivity for asteroid detection. The camera has a 4096 x 4096 pixel detector covering a $22.3'$ x $22.3'$ field of view (FoV), with a pixel size of $0.33''$, enabling accurate astrometry over a relatively wide field. Since the same sky strip is revisited at the same sidereal times each year, data from four observing cycles were processed uniformly with the optimised \texttt{PyLMT}, allowing consistent detection and multi-epoch analysis of minor planets across years.

\begin{figure}[htbp]
\centering
\includegraphics[width=0.8\textwidth]{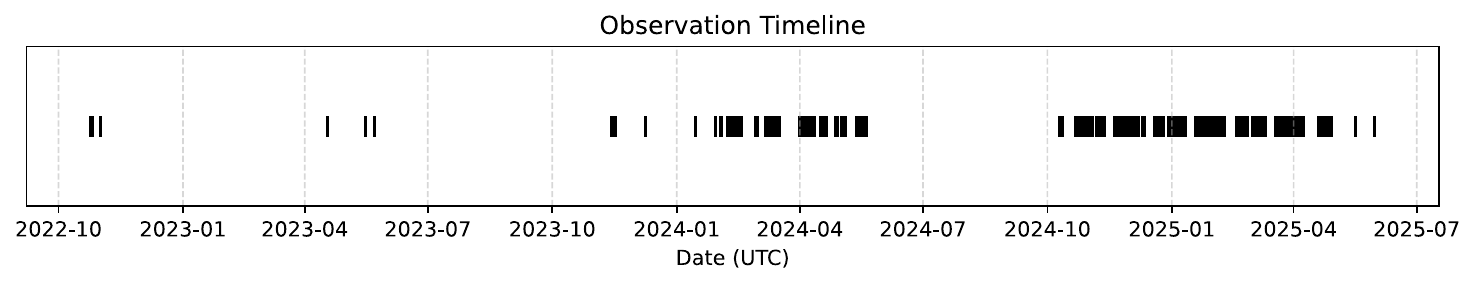}
\begin{minipage}{12cm}
\centering
\caption{Temporal distribution of observations over four observing cycles from 24 October 2022 to 30 May 2025, derived from the timestamps of images containing at least one detected transient or moving-object candidate.}
\label{fig:timeline}
\end{minipage}
\end{figure}

To illustrate the temporal sampling of the survey, we constructed an observation timeline from the timestamps of images containing at least one detected transient or moving-object candidate. The resulting timeline over the four cycles (24 October 2022 - 30 May 2025) is shown in Figure \ref{fig:timeline}. Gaps in the timeline are primarily due to the annual monsoon shutdown of telescope operations (approximately 15 June - 1 October) and occasional maintenance periods; other gaps were not individually checked.

\section{Methods of detection of minor planets}

Since the ILMT observes the same strip of sky each night with only a time shift of about 3 min 56 s, corresponding to the sidereal drift, minor planets can be detected through image subtraction between nights. This approach allows the use of a wide selection of reference images for the science frames, provided the reference image has better seeing \citep{pranshu2025pylmt}. The method can thus be automated and applied uniformly across the entire observed strip. We implemented this strategy using \texttt{PyLMT}, which reliably detects transient sources and discriminates minor planet candidates with good accuracy \citep{pranshu2025pylmt}.

\begin{figure}[H]
\centering
\includegraphics[width=0.8\textwidth]{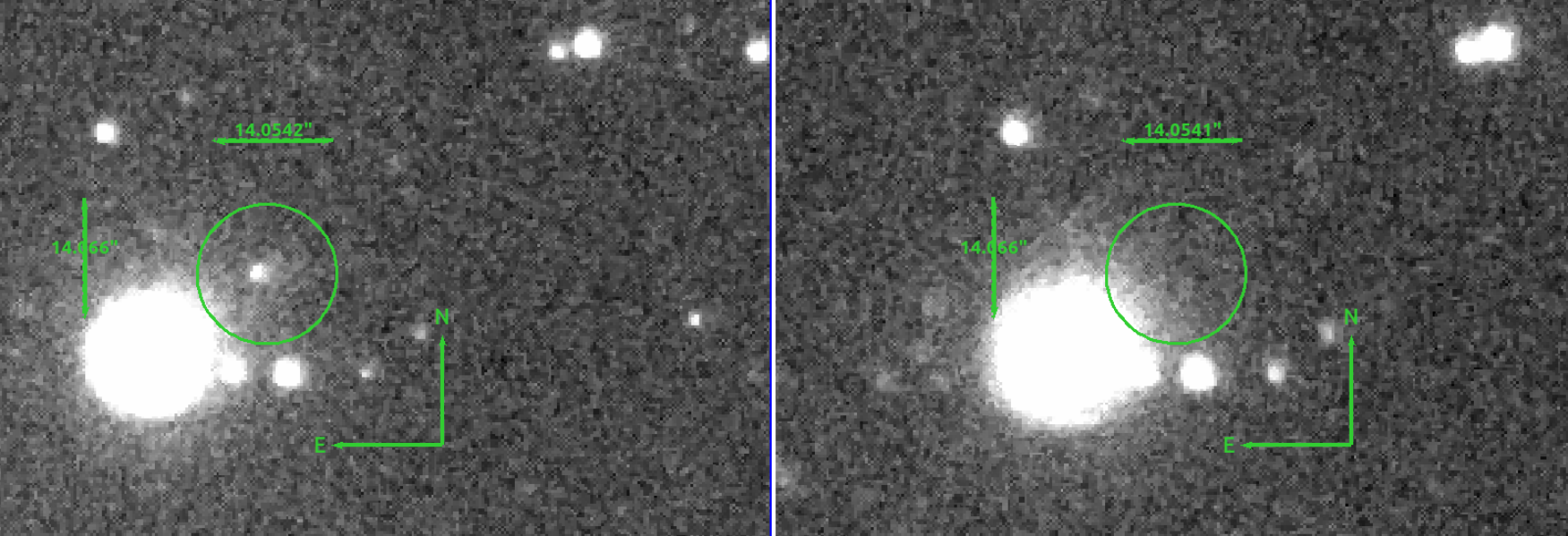}
\begin{minipage}{12cm}
\centering
\caption{Visual confirmation of the comet 498P/LINEAR through visual comparison of two frames obtained on 08 November 2024 (left) and 11 November 2024 (right). A point-spread function (PSF) consistent with the comet is visible in the 08 November 2024 frame, while no corresponding source is detected at the same position in the 11 November 2024 frame.}
\label{fig:498PLINEAR}
\end{minipage}
\end{figure}

After the application of \texttt{PyLMT} to our dataset, we performed a visual inspection of its output. An example of the visual confirmation is shown in Figure \ref{fig:498PLINEAR} for the comet 498P/LINEAR. Probable minor planets were selected by removing false positives, primarily artefacts and photometric transients identified via cross-matching with existing catalogues. In total, \texttt{PyLMT} identified about $41923$ transients, of which $21505$ were classified and visually confirmed as minor planet candidates.

\subsection{The ILMT's global efficiency in detecting asteroids}

To assess the astrometric accuracy of our measurements and validate the identifications, we compared the observed positions of the visually confirmed minor planet candidates with the MPC-predicted positions of the nearest objects at the epoch of observation, using the Minor Planet Checker (MPChecker) (https://www.minorplanetcenter.net/cgi-bin/checkmp.cgi). We define the measured separation as the angular distance between the observed position and the corresponding MPC-predicted position:

$\Delta r = \arccos \big[ \sin(\mathrm{Dec}_{\rm MPC}) \cdot \sin(\mathrm{Dec}_{\rm ILMT}) + \cos(\mathrm{Dec}_{\rm MPC}) \cdot \cos(\mathrm{Dec}_{\rm ILMT}) \cdot \cos(\mathrm{RA}_{\rm ILMT} - \mathrm{RA}_{\rm MPC}) \big]$,\\

where $\mathrm{RA}_{\rm MPC}, \mathrm{Dec}_{\rm MPC}$ are the predicted right ascension and declination returned by MPChecker, and $\mathrm{RA}_{\rm ILMT}, \mathrm{Dec}_{\rm ILMT}$ are the ILMT-measured positions,\\

while the predicted separation corresponds to the angular separation expected from the MPChecker uncertainties:

\begin{equation}
\mathrm{Predicted}(\Delta \mathrm{r}) = \mathrm{-24 \, \Delta t \, \sqrt{v_{\mathrm{RAcosDec,MPC}}^2 + v_{\mathrm{Dec,MPC}}^2}}
\end{equation}

\begin{figure}[H]
\centering
\includegraphics[width=0.8\textwidth]{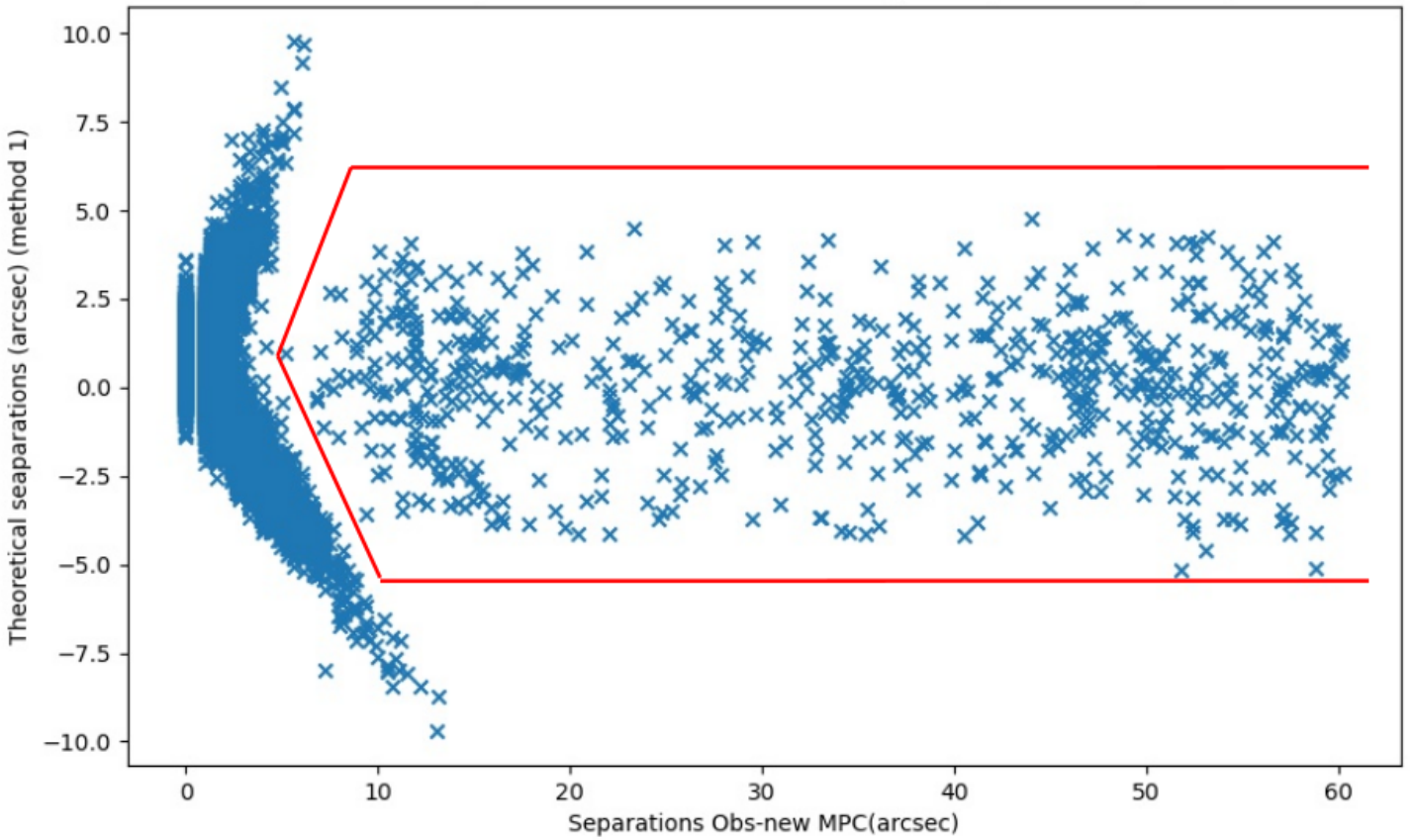}
\begin{minipage}{12cm}
\centering
\caption{Comparison between ILMT-MPC predicted and measured separations. Points enclosed within the red region correspond to sources that are not considered robust identifications and are therefore excluded from our present analysis. All other points are used in the subsequent astrometric and photometric analyses.}
\label{fig:theor_sep_vs_sep}
\end{minipage}
\end{figure}

$\Delta \mathrm t$ being the time difference between the MPChecker epoch and the observation time, expressed in days, because the observation epoch is initially computed with a precision of five decimal places and subsequently rounded to $0.01$ days, as imposed by the MPChecker input format:

\begin{equation}
\Delta \mathrm{t} =
\mathrm{datetime}_{\rm MPCForm}
- \mathrm{datetime}.
\end{equation}

and $v_{\mathrm{RAcosDec,MPC}}$ and $v_{\mathrm{Dec,MPC}}$ are the MPC-predicted components of the apparent motion of the minor planet candidate in right ascension and declination, respectively. The right-ascension component includes the $\cos(\mathrm{Dec})$ correction,

Figure~\ref{fig:theor_sep_vs_sep} shows the predicted separation as a function of the measured separation. While a linear trend is expected if the identifications are reliable, such a correlation is only observed up to measured separations of about $\sim 10''$. Even within this range, some points clearly deviate from the linear relation. These outliers, for which no visually convincing linear correlation is established, are highlighted in Figure~\ref{fig:theor_sep_vs_sep}. They are excluded from the present analysis, and their investigation is deferred to future work, as they cannot be considered robustly confirmed at this stage.

\section{Astrometric validation}
To retain only minor planet candidates located in the linear correlation regime between predicted and measured MPC-ILMT separations and to further constrain the identifications, we applied an additional astrometric criterion based on the residual angular separation. We define the residual angular separation as:

\begin{equation}
\begin{array}{lcl}
\mathrm{Residual}(\Delta r) &=&
\sqrt{(\mathrm{Residual}(\Delta \mathrm{RAcosDec}))^2
+ (\mathrm{Residual}(\Delta \mathrm{Dec}))^2} \\[6pt]

\mathrm{Residual}(\Delta \mathrm{RAcosDec}) &=&
\Delta \mathrm{RAcosDec}
- \mathrm{Predicted}(\Delta \mathrm{RAcosDec}) \\

\mathrm{Residual}(\Delta \mathrm{Dec}) &=&
\Delta \mathrm{Dec}
- \mathrm{Predicted}(\Delta \mathrm{Dec})
\end{array}
\end{equation}

with

\begin{equation}
\begin{array}{lcl}
\Delta \mathrm{RAcosDec} &=&
(\mathrm{RA}_{\rm ILMT} - \mathrm{RA}_{\rm MPC})
\cos(\mathrm{Dec}_{\rm MPC}) \\
\Delta \mathrm{Dec} &=& \mathrm{Dec}_{\rm ILMT} - \mathrm{Dec}_{\rm MPC}
\end{array}
\end{equation}

and

\begin{equation}
\begin{array}{lcl}
\mathrm{Predicted}(\Delta \mathrm{RAcosDec})
&=& \mathrm{-24 \, \Delta t \, v_{\mathrm{RAcosDec,MPC}}} \\
\mathrm{Predicted}(\Delta \mathrm{Dec})
&=& \mathrm{-24 \, \Delta t \, v_{\mathrm{Dec,MPC}}}
\end{array}
\end{equation}

We consider objects as confirmed minor planets when the residual angular separation is $\le 3''$. We adopt this threshold in view of the current astrometric precision of the \texttt{PyLMT}'s output. Since \texttt{PyLMT} does not yet provide centroid coordinates but instead reports the position of the first detected pixel of the point-spread function (PSF), the measured positions carry an uncertainty of approximately  $2''$, corresponding to the PSF size. An additional $1''$ accounts for uncertainties in the ephemeris calculation and coordinate transformations, leading to a total tolerance of $3''$. 

We therefore retain only minor planets with residuals $\le 3''$ relative to the MPC-predicted positions for the subsequent analysis. This selection ensures that only sources with reliable astrometry are retained for further analysis. We then characterise this confirmed sample through the complementary studies described below.

\subsection{Astrometric accuracy}

In total, we identified $19903$ asteroids, corresponding to $9007$ unique objects, and four comet observations: 498P/LINEAR, 492P/LINEAR, 28P/Neujmin, and 384P/Kowalski. A summary of these detected comets is provided in Table \ref{tab:comets}. They were excluded from the subsequent statistical analysis, which considers only confirmed asteroids.

\begin{figure}
\centering
\includegraphics[height=0.3\textheight, width=0.8\textwidth]{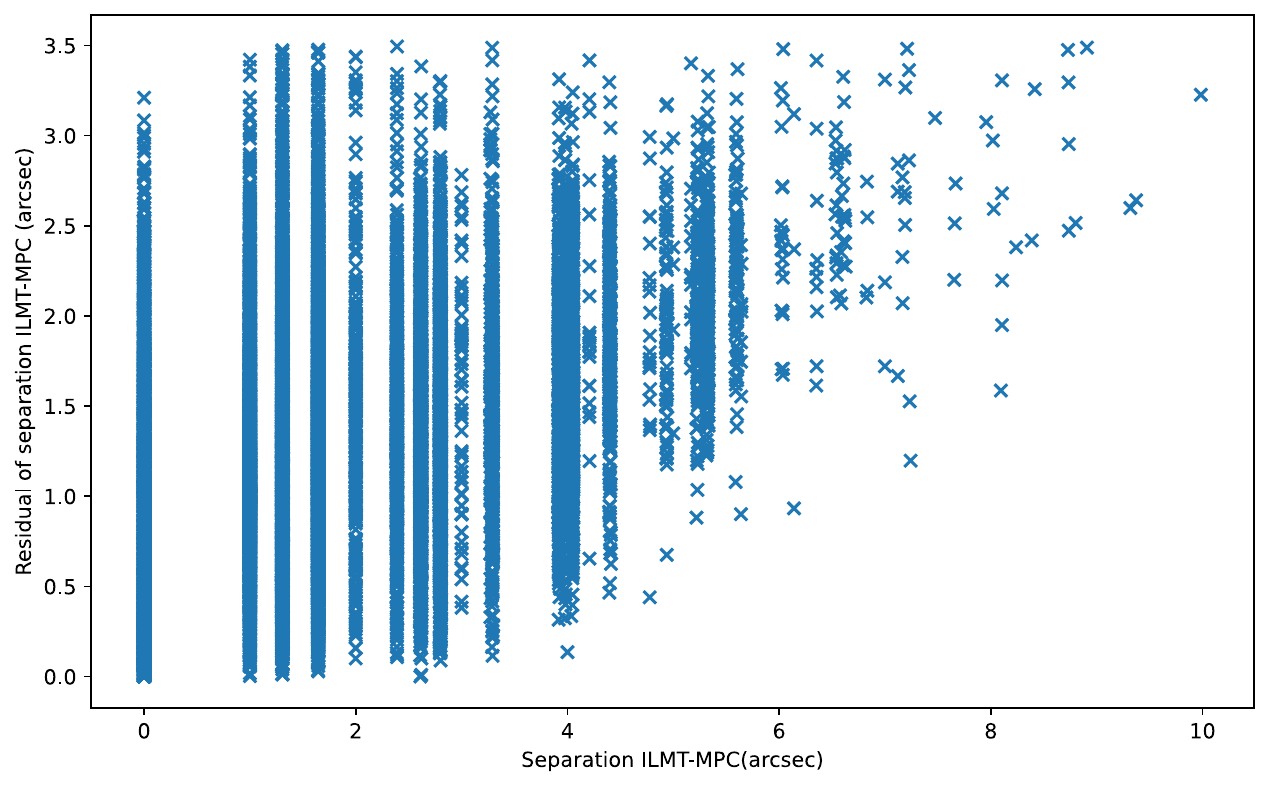}
\begin{minipage}{12cm}
\centering
\caption{Residual angular separation versus MPC-predicted separation for confirmed asteroids. Measured separations up to $\sim 10''$ can be accounted for when applying the $3''$ residual threshold, illustrating that the pipeline, visual confirmation, and astrometric validation reliably identify confirmed objects.}
\label{fig:resid_sep_vs_sep}
\end{minipage}
\end{figure}

To assess the astrometric consistency of our identifications, we plotted the residual separation versus the predicted separation (Figure \ref{fig:resid_sep_vs_sep}) for the confirmed asteroids. The distribution shows that, even for measured separations up to $\sim 10''$, applying the $3''$ residual threshold in combination with the \texttt{PyLMT} and visual confirmation yields a confirmed sample with reliable astrometry.

\begin{table}[htbp]
\centering
\begin{minipage}{160mm}  
\centering
\caption{Comets detected in the ILMT's Cycles 1-4. The table reports the observed positions, MPC designations, predicted $V$ magnitudes, and ILMT-MPC angular separations.}
\label{tab:comets}
\end{minipage}
\begin{tabular}{ccccc}
\hline
Equatorial coordinates (J2000) & Observation date & Designation & V & Separation ($''$)\\
\hline
(07h 15m 57.5s, 29d 21m 15s) & 08-11-2024 & 498P/LINEAR & 15.3 & 1.6\\
(09h 44m 00.9s, 29d 26m 39s) & 03-12-2024 & 492P/LINEAR & 16.2 & 1.3\\
(08h 48m 25.6s, 29d 21m 01s) & 16-11-2023 & 28P/Neujmin & 12.9 & 1\\
(09h 29m 54.3s, 29d 37m 16s) & 25-12-2024 & 384P/Kowalski & 18.9 & 2.8\\

\hline
\end{tabular}
\end{table}

\subsection{Spatial distribution in the Ecliptic}

\begin{figure}[H]
\centering
\begin{minipage}{5.04cm}
    \textbf{(a)}\\
    \includegraphics[width=\linewidth]{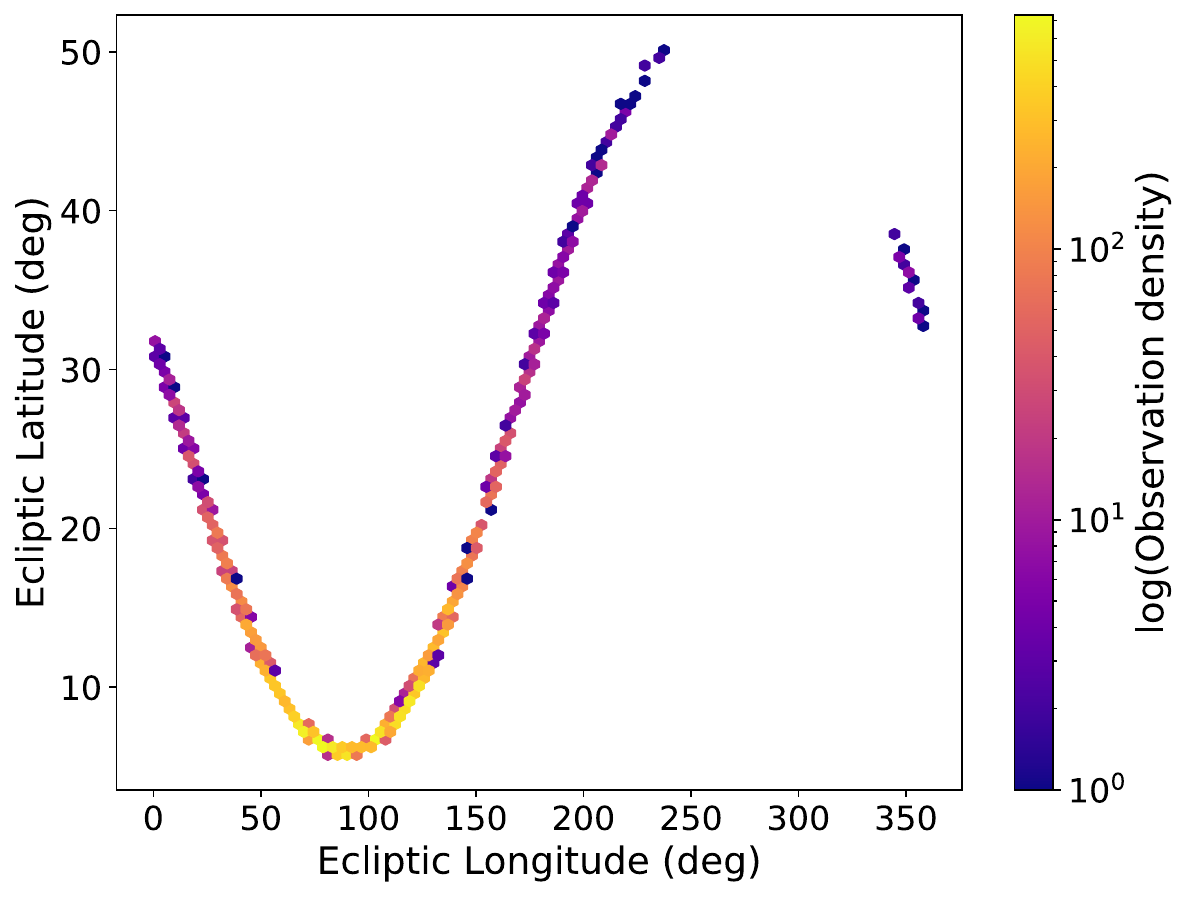}
\end{minipage}
\begin{minipage}{6.72cm}
    \textbf{(b)}\\
    \includegraphics[width=\linewidth]{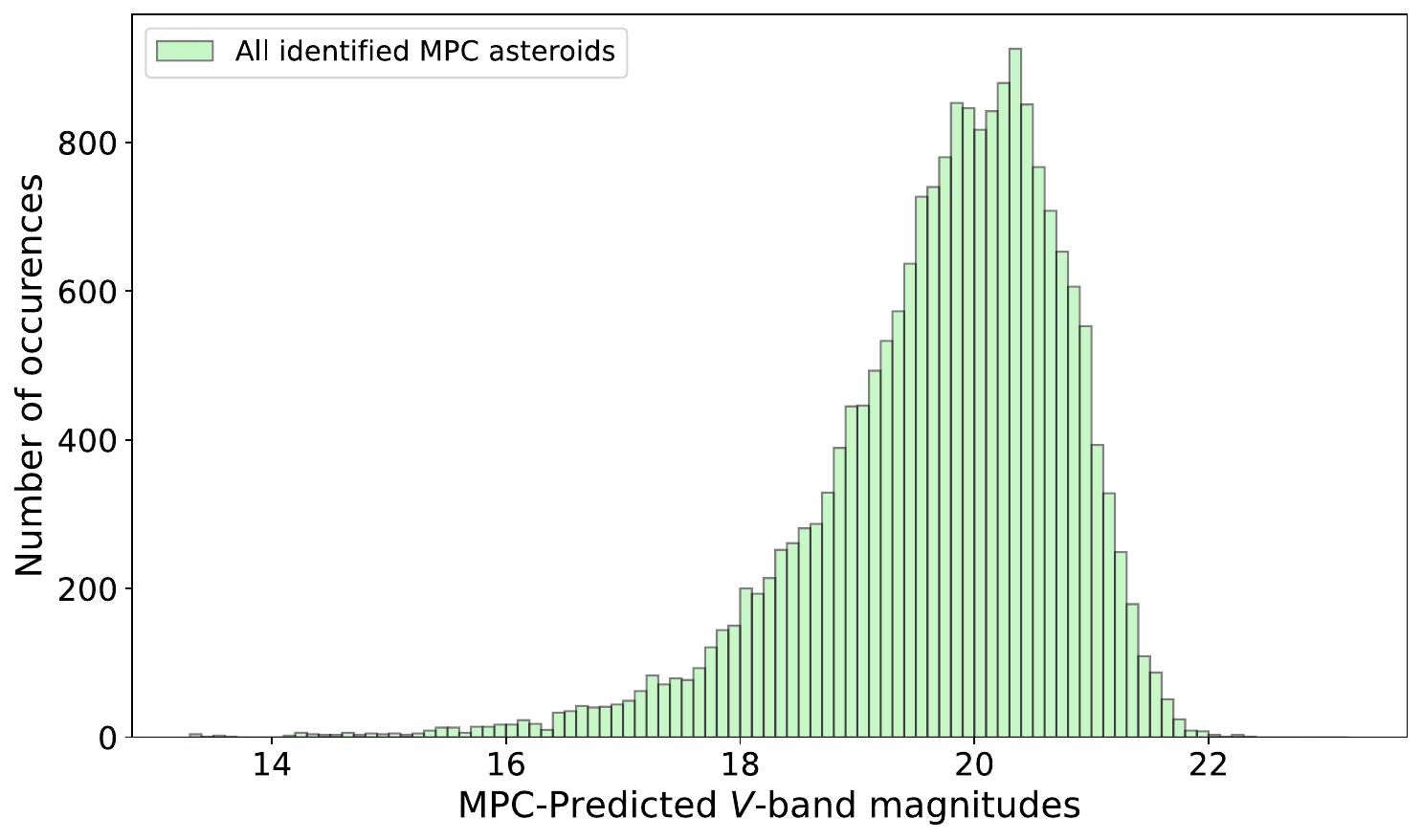}
\end{minipage}
\begin{minipage}{12cm}
\caption{(a) Hexagonal density map of asteroid detections in ecliptic longitude and latitude. Colors indicate the logarithmic number density of objects per hexagonal bin. (b) Distribution of MPC-predicted V magnitudes for the astrometrically confirmed asteroids.
}
\end{minipage}
\label{fig:density_in_ecliptic_vmags_less3as}
\end{figure}

We examined the distribution of the confirmed asteroids as a function of their ecliptic coordinates. Figure \ref{fig:density_in_ecliptic_vmags_less3as}(a) shows the density of objects as a function of ecliptic longitude and latitude. This analysis highlights the concentration of main-belt asteroids near the ecliptic plane and provides a global view of the spatial coverage of the survey. As expected, most minor planets are detected at low to moderate ecliptic latitudes $\beta$, with $59\%$ of detections occuring at $5.7 ^\circ \le \beta \le 10^\circ $ and $90\%$ at $5.7 ^\circ \le \beta \le 18^\circ$. The relative deficit of detections between ecliptic longitudes $235^\circ$ and $345^\circ$ can be attributed to multiple factors. First, this longitude range in ILMT's survey corresponds to high ecliptic latitudes ($38^\circ \le \beta \le 50^\circ$), where asteroids are intrinsically less numerous. Second, it coincides with the observation period from 27 April 20:39:13 UT1 (Universal Time 1) to 31 October 14:01:45 UT1, corresponding to the interval in Indian Standard Time, IST $\approx 19:30$ - $02:00$. In other words, the observed gap results from the absence of observations during the monsoon season, together with reduced coverage in the early night hours in the adjacent periods, caused by seasonal weather conditions.

\section{Photometric properties}

To characterise the brightness distribution of the astrometrically confirmed sample, we constructed a histogram of MPC-predicted V magnitudes shown in Figure \ref{fig:density_in_ecliptic_vmags_less3as}(b). The confirmed asteroids span MPC-predicted magnitudes V$\approx 13.3$---$23.2$, with a peak near V$\approx 20.3$, and the number of detected objects decreases to roughly half the peak value at V$\approx 21$. This distribution is broadly similar to that found in a previous study \citep{filali2026asteros} and provides an empirical estimate of the effective detection depth.

\section{Object classification and population statistics}

We further classify the detected bodies and determine the number of NEOs and PHAs present in our sample. Detecting and characterizing these populations is important for refining the orbital solutions of individual objects and for constraining the statistical properties of small-body populations within the ILMT's FoV, while contributing to ongoing efforts to identify and monitor potentially hazardous objects. We summarize the astrometric and photometric properties of detected NEOs and PHAs in Table \ref{special_mp_table}. Furthermore, many of the identified asteroids were observed on multiple nights, providing repeated sampling of individual objects. This makes it possible, in future work, to construct phase-folded light curves and study their rotation through phase-folded photometry.

\section{Conclusion}

The ILMT successfully observed $21505$ minor planet candidates detected with \texttt{PyLMT} and validated through visual inspection. To these, we applied an astrometric validation threshold of $3''$, justified by internal and numerical astrometric uncertainties, thus confirming $19903$ catalogued asteroids representing $9007$ distinct objects observed on multiple nights, and $4$ comets.

We examined the residual angular separation as a function of measured-MPC positional offset, and found that separations up to $\sim 10''$ are consistent with the adopted $3''$ residual threshold, providing guidance for future automation of MPC cross-matching.

We performed a preliminary analysis of the distribution of the confirmed asteroids in ecliptic coordinates, showing higher densities at low to intermediate ecliptic latitudes. We also built the histogram of MPC-predicted V magnitudes for the confirmed asteroids. Magnitudes range from $\sim 13.3$ to $\sim 23.2$, with a peak near V$\approx 20.3$. The number of detected objects decreases to about half the peak value at V $\approx  21$. These results provide an empirical estimate of the survey's effective detection depth.

Among the confirmed asteroids, more than $50 \%$ were observed during multiple nights, and 19 were observed during at least $15$ nights. Such repeated sampling will enable the construction of light curves and the future determination of their rotational periods.

In total, $18$ observations correspond to NEOs and PHAs, representing $14$ distinct objects. In addition, $1598$ non-confirmed minor planet candidates remain to be carefully analysed; some of these may correspond to previously unknown asteroids or other transient phenomena.

\clearpage
\appendix
\addcontentsline{toc}{section}{Appendix}
\section{Supplementary Table}

\begin{table}[H]
\centering
\begin{minipage}{140mm}  
\centering
\caption{NEOs and PHAs detected in ILMT's Cycles 1-4. Designations follow the MPC nomenclature. Right ascension and declination correspond to the measured astrometric positions (J2000) at the epoch of detection. V (MPC) denotes the MPC-predicted visual magnitude at the same epoch. The class (NEO or PHA) is adopted from the MPC classification.}
\label{special_mp_table}
\end{minipage}
\begin{tabular}{ccccc}
\hline
Designation & RA (J2000.0) & Dec (J2000.0) & V (MPC) & Class\\
\hline
(380981) 2006 SU131 & 12h 15m 58.50s & 29d 33m 12.0s & 19 & PHA\\
(613986) 2008 JG & 08h 23m 29.90s & 29d 20m 20.0s & 19 & PHA\\
(762379) 2011 CG2 & 05h 59m 20.10s & 29d 23m 56.0s & 20.1 & PHA\\
(185851) 2000 DP107 & 01h 58m 44.60s & 29d 13m 35.0s & 20.2 & PHA\\
2001 QB34 & 04h 53m 46.20s & 29d 23m 52.0s & 20.6 & NEO\\
2024 UJ13 & 01h 36m 25.00s & 29d 16m 37.0s & 21.4 & NEO\\
2024 RZ6 & 00h 54m 50.20s & 29d 11m 37.0s & 21 & NEO\\
(887) Alinda & 11h 04m 57.00s & 29d 39m 11.0s & 14.9 & NEO\\
(887) Alinda & 11h 05m 39.40s & 29d 21m 09.0s & 15 & NEO\\
(40267) 1999 GJ4 & 11h 09m 15.00s & 29d 24m 45.0s & 19 & NEO\\
(25916) 2001 CP44 & 08h 43m 02.00s & 29d 36m 48.0s & 19.2 & NEO\\
(25916) 2001 CP44 & 08h 44m 01.10s & 29d 31m 11.0s & 19.2 & NEO\\
(25916) 2001 CP44 & 08h 44m 59.90s & 29d 25m 29.0s & 19.2 & NEO\\
(144861) 2004 LA12 & 14h 58m 11.80s & 29d 21m 03.0s & 20.1 & NEO\\
(155334) 2006 DZ169 & 03h 54m 43.50s & 29d 19m 40.0s & 20.4 & NEO\\
(96189) Pygmalion & 05h 44m 28.90s & 29d 18m 38.0s & 18.7 & NEO\\
(420187) 2011 GA55 & 02h 07m 20.60s & 29d 11m 19.0s & 20.9 & NEO\\
(420187) 2011 GA55 & 02h 08m 48.80s & 29d 18m 51.0s & 20.9 & NEO\\

\hline
\end{tabular}
\end{table}

\begin{acknowledgments}

This work is a result of the Belgo-Indian Network for Astronomy and Astrophysics (BINA), which promotes astronomical collaborations between Indian and Belgian partners and received financial support from the International Division of the Department of Science and Technology (DST, Government of India) and the Belgian Federal Science Policy Office (BELSPO, Government of Belgium) during the period 2016-2023 through different bilateral projects.
The 4-m International Liquid Mirror Telescope (ILMT) project results from a collaboration between the Institute of Astrophysics and Geophysics (University of Liège, Belgium), the Universities of British Columbia, Laval, Montreal, Toronto, Victoria and York, and Aryabhatta Research Institute of observational sciencES (ARIES, India). KP acknowledges the support
from the Erasmus+ Programme of the European Union for a research visit to the Institute of Astrophysics and Geophysics, University of Liège, Belgium. KM acknowledges the support from the BRICS grant DST/ICD/BRICS/Call-5/CoNMuTraMO/2023 (G) funded by the DST, India. JS wishes to thank Service Public Wallonie, F.R.S.-FNRS (Belgium) and the University of Liège, Belgium, for funding the construction of the ILMT. ARIES thanks the Department of Science and Technology (DST), Govt. of India, for the realisation of the project. JS and KM acknowledge the assistance received from the Anusandhan National Research Foundation
(ANRF, SERB- 762 VAJRA Faculty Scheme, India). PH acknowledges financial support from the Natural Sciences and Engineering Research Council of Canada, RGPIN-2019-04369.

\end{acknowledgments}

\begin{furtherinformation}

\begin{orcids}

  \orcid{0009-0009-3108-3789}{Sara}{FILALI}
  \orcid{0000-0002-7005-1976}{Jean}{SURDEJ}

\end{orcids}

\begin{authorcontributions}

This work results from a long-term collaboration to which all authors have made significant contributions.

\end{authorcontributions}

\begin{conflictsofinterest}

The authors declare no conflict of interest.

\end{conflictsofinterest}

\end{furtherinformation}

\bibliographystyle{bullsrsl-en}

\bibliography{extra.bib}

\end{document}